\documentstyle[prd,aps]{revtex}

\newcommand{\td}{r^2}

\newcommand{\rhop}{\rho_+}
\newcommand{\rhom}{\rho_-}
\newcommand{\rhod}{\rho^2}

\newcommand{\srho}{\sqrt{\rho^2-1}}

\newcommand{\be}{\begin{equation}}
\newcommand{\ee}{\end{equation}}
\newcommand{\bea}{\begin{eqnarray}}
\newcommand{\eea}{\end{eqnarray}}

\newcommand{\bei}{\begin{itemize}}
\newcommand{\eei}{\end{itemize}}
\newcommand{\sla}[1]{#1\!\!\!\slash}

\newcommand{\xp}{x^\prime}

\newcommand{\Ep}{{E^\prime}}
\newcommand{\El}{{E_l}}
\newcommand{\Mb}{M_{\Lambda_b}}
\newcommand{\Mc}{M_{\Lambda_c}}

\begin{document}

\author{Hsien-Hung Shih\footnote{hhshih@phys.sinica.edu.tw} and
Shih-Chang Lee\footnote{phsclee@ccvax.sinica.edu.tw}
\\
{\small Institute of Physics, Academia Sinica, Taipei, Taiwan 105,
Republic of China}
\\
Hsiang-nan Li\footnote{hnli@mail.ncku.edu.tw}
\\
{\small Department of Physics, National Cheng-Kung University,
Tainan, Taiwan 701, Republic of China} 
\\
{\small Theory Group, KEK, Tsukuba, Ibaraki 305, Japan}}

\title{Polarized $\Lambda_b \to \Lambda_c l{\bar\nu}$ decay in
perturbative QCD}

\date{\today}

\maketitle

\begin{abstract}
We study the polarized $\Lambda_b$ baryon semileptonic decay 
$\Lambda_b \to \Lambda_cl{\bar\nu}$ using perturbative QCD factorization 
theorem. Employing the heavy baryon transition form factors calculated 
in our previous work, we predict the asymmetry parameter
$\alpha({\hat e})\equiv (N_+-N_-)/(N_++N_-)=-0.31{\bf P}\cdot{\hat e}$,
where ${\bf P}$ is the $\Lambda_b$ baryon polarization, ${\hat e}$ a unit
vector perpendicular to the $\Lambda_b$ baryon momentum, $N_+$ ($N_-$)
the number of leptons whose momenta possess positive (negative)
components along ${\hat e}$. This result is useful for extracting the
transverse polarization of a $\Lambda_b$ baryon from data of exclusive
semileptonic decays.

\end{abstract}
\vskip 0.5cm

\section{Introduction}

Recently, we have developed perturbative QCD (PQCD) factorization theorem 
for exclusive heavy baryon decays \cite{SLL,SLL2}. In this approach 
transition form factors are expressed as convolutions of hard heavy 
quark decay amplitudes with heavy baryon wave functions. The former are 
calculable in perturbation theory in the kinematic region where large 
momentum transfer is involved. The latter, absorbing nonperturbative 
dynamics of decay processes, must be obtained by means outside the PQCD 
regime. Large logarithmic corrections are organized by Sudakov 
resummation and renormalization-group equations to improve
perturbative expansions. Since wave functions are universal, they can be 
determined once for all, and then employed to make predictions for other 
modes containing the same heavy baryons. With this prescription for 
nonperturbative wave functions, PQCD factorization theorem possesses 
predictive power.  

In this paper we shall apply the above PQCD formalism to polarized
$\Lambda_b$ baryon decays. It is well known that hyperons produced in 
hadronic machines through strong interaction are usually polarized. 
However, the mechanism responsible for such a polarization is still 
unclear. Hence, it is interesting to investigate if similar phenomena 
happen for heavy baryons produced in proton-proton and proton-anti-proton 
collisions at various energies. Since $b$-baryon events in hadronic 
productions are often selected by the presence of high energy leptons, we 
shall focus on studying the $\Lambda_b$ baryon polarization through the 
semileptonic decays. The longitudinal polarization of a $\Lambda_b$ baryon 
produced from $Z^0$ decays \cite{ALEPH,OPAL} could be determined by 
measuring the ratio of the average electron (muon) energy to the average 
neutrino energy in the inclusive semileptonic decays \cite{BR,DTKP}. This 
ratio, obtained in the laboratory frame, is then related to the ratio of 
the corresponding energies in the rest frame of the $\Lambda_b$ baryon. 

When heavy baryons are produced via strong interaction, they could possess 
transverse polarization perpendicular to baryon momenta. In the present 
work we shall concentrate on the determination of the transverse 
polarization of a $\Lambda_b$ baryon from the angular distribution of 
leptons produced in the $\Lambda_b \to \Lambda_c l{\bar\nu}$ decay. For 
semileptonic decays, it is usually not possible to reconstruct fully the 
momentum of the parent baryon. In order to determine the heavy baryon 
polarization through semileptonic events, we propose to measure the 
asymmetry parameter $\alpha\equiv (N_+-N_-)/(N_++N_-)$, where $N_+$ 
($N_-$) is the lepton number emitted above and below a plane that 
contains the $\Lambda_b$ baryon momentum. This quantity is invariant under 
a longitudinal boost and can be simply evaluated in the rest frame of the 
$\Lambda_b$ baryon. Using the $\Lambda_b \rightarrow\Lambda_c$ transition 
form factors obtained from PQCD factorization theorem \cite{SLL2}, we 
derive the relation between $\alpha$ and the transverse polarization 
${\bf P}\cdot {\hat e}$ of the $\Lambda_b$ baryon, $\alpha=-0.31 {\bf P}
\cdot {\hat e}$, where ${\hat e}$ is a unit vector perpendicular to the 
$\Lambda_b$ baryon momentum. Comparing this relation with experimental 
data of $\alpha$, one may extract the $\Lambda_b$ baryon transverse 
polarization. 

It has been shown that PQCD analyses for $\Lambda_b\to \Lambda_c$ 
transition form factors are reliable at the maximal recoil of the 
$\Lambda_c$ baryon \cite{SLL2}. The percentages of the full contribution
to the form factors, that arise from the short-distance region with
$\alpha_s/\pi<0.2$-0.5, are listed in Table I. It is observed that
for the velocity transfer $\rho \sim 1.4$, about 60\% of the full
contribution comes from the region with $\alpha_s/\pi<0.3$. Though the
results are not completely perturbative, it makes sense to estimate
the form factors at the maximal recoil using PQCD, and to give predictions
that can be tested by future experiments.

Extrapolating PQCD results to small $\rho$ under the requirement of heavy
quark symmetry (HQS) \cite{IW}, we have determined the behaviors of
the form factors in the whole range of the velocity transfer. This
extrapolation leads to the reasonable branching ratio
$B(\Lambda_b \to \Lambda_c l{\bar\nu})=2\sim 3\%$, which is consistent
with the experimental upper bound of the branching ratio from the data
$B(\Lambda_b\to \Lambda_cl{\bar\nu}+X)=(8.27\pm 3.38)\%$ \cite{Data}. The
$\Lambda_b\to\Lambda_c$ transition form factors have been evaluated by
means of overlap integrals of infinite-momentum-frame wave functions
\cite{XHG,KKK}, relativistic quark models \cite{IKL1}, Bethe-Salpeter
equations \cite{IKL2} and QCD sum rules \cite{DFN,CNN,Dai2}. Most of the
analyses led to the branching ratios about or below 6\%. Our result is
close to the result $(3.4\pm 0.6)\%$ derived in \cite{DFN}.

\section{Asymmetry parameter}

We first define the $\Lambda_b\to \Lambda_c$ transition form factors, and
refer their explicit factorization formulas to \cite{SLL2}. The amplitude
for the semileptonic decay $\Lambda_b\to \Lambda_cl\bar{\nu}$ is written
as
\bea
{\cal M} = \frac{G_F}{\sqrt{2}}V_{cb} \; \bar{l}\gamma^{\mu} 
(1-\gamma_5) \nu_l \; \langle \Lambda_c(p^\prime)| \bar{c}\gamma_{\mu}
(1-\gamma_5)b |\Lambda_b(p)\rangle \;,
\label{mm}
\eea
where $G_F$ is the Fermi coupling constant, $V_{cb}$ the
Cabibbo-Kobayashi-Maskawa (CKM) matrix element, $p$ and $p^\prime$ the
$\Lambda_b$ and $\Lambda_c$ baryon momenta, respectively. QCD dynamics
is contained in the hadronic matrix element
\bea
{\cal H}_{\mu} &\equiv& \langle \Lambda_c(p^\prime)|\bar{c}\gamma_{\mu} 
(1-\gamma_5)b| \Lambda_b(p)\rangle\;,
\nonumber\\
&=&\bar{\Lambda}_c(p') [ f_1(q^2)\gamma^\mu
-if_2(q^2)\sigma^{\mu\nu}q_\nu+f_3(q^2)q^\mu ]\Lambda_b(p)
\nonumber \\
& & +\bar{\Lambda}_c(p')[ g_1(q^2)\gamma^\mu
-ig_2(q^2)\sigma^{\mu\nu}q_\nu
+g_3(q^2)q^\mu ]\gamma_5\Lambda_b(p)\;.
\eea
In the second line ${\cal H}_\mu$ has been expressed in terms of six
form factors $f_i$ and $g_i$, where $\Lambda_b(p)$ and $\Lambda_c(p')$ are
the $\Lambda_b$ and $\Lambda_c$ baryon spinors, respectively, and the 
variable $q$ denotes $q=p-p'$. In the case of massless leptons with
$q_\mu \bar{l} \gamma^\mu(1-\gamma_5) \nu_l= 0$, $f_3$ and $g_3$ do not 
contribute. Since the contributions from $f_2$ and $g_2$ are small, we 
shall consider only $f_1$ and $g_1$ below.

The polarization density matrix for a $\Lambda_b$ baryon 
with polarization $P$ is written as 
\bea
\varrho_{KK^\prime} &=& \frac{1}{2} {P}^a
(\sigma_a)_{KK^\prime} \;,\;\;\;\;\sigma_0=I\;.
\label{pd}
\eea
For the purpose of normalization, we usually choose ${P}^0=1$.
The hadronic matrix element with the helicity of the initial- and 
final-state baryons specified is expressed as
\bea
{\cal H}_\mu(K,J) &=& \bar{\Lambda}_c(p^\prime,J)[f_1(\rho)\gamma_\mu
+g_1(\rho)\gamma_\mu\gamma_5]\Lambda_b(p,K)\;,
\eea
which leads to the covariant density matrix 
\bea
{\cal H}_{\mu\nu} &=& \sum_{K,K',J}{\cal H}_\mu(K,J)
\varrho_{KK^\prime} {\cal H}^*_\nu(K',J)\;.
\label{lpt}
\eea
The differential decay rate is written as
\bea
d\Gamma &=& \frac{1}{2 M_{\Lambda_b}}|{\cal M}_p|^2 
(2\pi)^4\delta^4(p-p^\prime-p_l-p_\nu)
\frac{d^3{\bf p}^\prime}{(2\pi)^3 2E^\prime}
\frac{d^3{\bf p}_l}{(2\pi)^3 2E_l}
\frac{d^3{\bf p}_\nu}{(2\pi)^3 2E_{\nu}}\;,
\label{pdr}
\eea
where $p_l$ ($p_\nu$) is the lepton (neutrino) momentum. The matrix 
element square is given by
\begin{equation}
|{\cal M}_p|^2 = \frac{G_F^2}{2}|V_{cb}|^2 {\cal H}_{\mu\nu} 
{\cal L}^{\mu\nu}\;,
\label{msq}
\end{equation} 
with the leptonic covariant tensor,
\begin{equation}
{\cal L}^{\mu\nu} = 8 \left[ p_l^\mu p_\nu^\nu+p_l^\nu p_\nu^\mu-
p_l \cdot p_\nu g^{\mu\nu} + 
i\epsilon^{\mu\nu\alpha\beta}{p_l}_\alpha {p_\nu}_{\beta} \right]\;.
\end{equation}

Define the velocity transfer $\rho$, which is related to $q^2$ via
\bea 
\rho = \frac{M_{\Lambda_b}^2+M_{\Lambda_c}^2-q^2}
{2M_{\Lambda_b} M_{\Lambda_c}} \;,
\label{rr}
\eea
with $M_{\Lambda_b}$ and $M_{\Lambda_c}$ being the $\Lambda_b$ and
$\Lambda_c$ baryon masses, respectively.
Performing the phase space integrations straightforwardly, Eq.~(\ref{pdr}) 
reduces to 
\bea
\frac{d^4 \Gamma}{dE'dE_ld\cos{\theta_c}d\Omega_l} &=&
\frac{1}{256 \pi^5 M_{\Lambda_b}}
\frac{1}{\sin\theta_c \sin\theta_l \sin{\bar\phi}}
|{\cal M}_p|^2\;, 
\label{dr2}
\eea
where $E'=\rho \Mc$ ($E_l$) is the $\Lambda_c$ baryon (lepton)
energy, $\theta_c$ ($\theta_l$) the polar angle 
of the $\Lambda_c$ baryon momentum ${\bf p}'$ (the lepton momentum 
${\bf p}_l$) against the $\Lambda_b$ baryon polarization, $\phi_l$ 
the azimuthal angle of ${\bf p}_l$, and $d\Omega_l = d\cos\theta_l d\phi_l$
the solid angle. The azimuthal angle $\bar{\phi}$ and the angle 
$\bar\theta$ between ${\bf p}'$ and ${\bf p}_l$ satisfy the relations
\bea
\cos\bar{\phi} &=& \frac{1}{\sin\theta_c \sin\theta_l}
[ \cos{\bar\theta}-\cos\theta_c \cos\theta_l ]\;,
\label{azi}\\
\cos{\bar\theta} &=& 
\frac{\Mb^2+\Mc^2-2\Mb\Ep-2\Mb\El+2\Ep\El}{2\sqrt{{\Ep}^2-\Mc^2}\El}\;.
\eea
The kinematic ranges of $\Ep$ and $\El$ are 
\bea
\Mc \le & \Ep & \le\frac{\Mb^2+\Mc^2}{2\Mb}\;,
\nonumber \\
\frac{1}{2} \left[\Mb-\Ep-\sqrt{{\Ep}^2-\Mc^2}\right]
\le & \El & \le\frac{1}{2} \left[ \Mb-\Ep+\sqrt{{\Ep}^2-\Mc^2}\right]\;.
\label{er}
\eea
Requiring $\cos^2{\bar\phi}\le 1$, it is easy to derive the ranges of 
$\cos\theta_c$ and $\cos\theta_l$ from Eq.~(\ref{azi}),
\bea
 &|\cos\theta_l|& \le 1 \;,
\nonumber \\
\cos(\theta_l+{\bar\theta})\le &\cos\theta_c& \le
\cos(\theta_l-{\bar\theta})\;.
\label{rt2}
\eea

The projection operator associated with a  spin-1/2 particle of mass $m$ 
is written as \cite{Lee}
\bea
u(p,K)\bar{u}(p,K^\prime) &=& \frac{1}{2}(m+\sla{p})[(I)_{K^\prime K}
       +\gamma_5\gamma_{\mu}n_i^{\mu}(\sigma^i)_{K^\prime K}]\;,
\label{po}
\eea
where $I$ is the $2\times 2$ unit matrix, $K$ and $K^\prime$ represent 
the helicities $\pm\frac{1}{2}$, and $\sigma^i$ are the Pauli matrices. 
In the rest frame of a $\Lambda_b$ baryon with 
$p=(M_{\Lambda_b},{\bf 0})$, we have the boost vectors $n^j_i=\delta^j_i$, 
$i,j=1,2,3$.
Using Eqs.~(\ref{pd}), (\ref{lpt}), and (\ref{po}), we obtain the 
expression of $|{\cal M}_p|^2$,
\bea
|{\cal M}_p|^2 &=&
8 \Mb G_F^2 |V_{cb}|^2 [ F_0(E^\prime,E_l)
+F_1(E^\prime,E_l) {\bf p}^\prime\cdot {\bf P} 
+F_2(E^\prime,E_l) {\bf p}_l \cdot {\bf P} ]\;,
\label{p2}
\eea
with the functions
\bea
F_0 &=& -(\Mb-\Ep-E_l) [\Mb^2+\Mc^2-2(\Ep+E_l) \Mb] (f_1-g_1)^2
\nonumber \\
    & &  +E_l (\Mb^2-\Mc^2-2 \Mb E_l) (f_1+g_1)^2
-\Mc (\Mb^2+\Mc^2-2 \Mb \Ep) (f_1^2-g_1^2)\;,
\nonumber \\
F_1 &=& [\Mb^2+\Mc^2-2(\Ep+E_l) \Mb] (f_1-g_1)^2
        +2 \Mc E_l (f_1^2-g_1^2)\;,
\nonumber \\
F_2 &=& [\Mb^2+\Mc^2-2(\Ep+E_l) \Mb] (f_1-g_1)^2
        - (\Mb^2-\Mc^2-2 \Mb E_l) (f_1+g_1)^2
\nonumber \\
    & & +2 \Mc (\Mb-\Ep) (f_1^2-g_1^2)\;,
\label{fun}
\eea
and the scalar products
\bea
{\bf p}^\prime\cdot {\bf P} = \sqrt{{E^\prime}^2-\Mc^2}
\;|{\bf P}| \cos\theta_c \;,\;\;\;\;
{\bf p}_l\cdot {\bf P} = E_l \; |{\bf P}|\cos\theta_l\;.
\eea
The first term $F_0$ is associated with the unpolarized $\Lambda_b$ 
baryon decay. Integrating out $\cos\theta_c$ according to the range in 
Eq.~(\ref{rt2}), Eq.~(\ref{dr2}) becomes
\bea
\frac{d^3\Gamma}{d\Ep dE_l d\Omega_l} &=& \frac{G_F^2}{16 \pi^4}
           |V_{cb}|^2 [ F_0+(\Mc\;\sqrt{\rho^2-1} \cos \bar{\theta} F_1 
           + E_l F_2)\;\hat{\bf p}_l \cdot{\bf P}  ] \;,
\label{dr3}
\eea
where $\hat{\bf p}_l={\bf p}_l/|{\bf p}_l|$ is the unit vector along 
the lepton momentum.

To obtain the decay rate, we need the information of the form factors
$f_1(\rho)$ and $g_1(\rho)$ in the whole range of $\rho$. $f_1$ and
$g_1$ have been evaluated in \cite{SLL2} by adopting the CKM matrix 
element $V_{cb}=0.04$, the masses $M_{\Lambda_b}=5.624$ GeV and 
$M_{\Lambda_c}=2.285$ GeV. As shown in Table I, the contribution from
the short-distance region with $\alpha_s/\pi<0.3$ becomes dominant
gradually when $\rho$ increases. The PQCD analysis can be regarded as
being self-consistent at the maximal $\rho\sim 1.4$, for which the
perturbative contribution amounts to about 60\% of the full contribution.
We then extrapolated the PQCD predictions to the small $\rho$ region under
the requirement of HQS \cite{IW}. Assuming the parametrization,
\begin{equation}
f_1(\rho)=\frac{c_f}{\rho^{\alpha_f}}\;,\;\;\;\;
g_1(\rho)=\frac{c_g}{\rho^{\alpha_g}}\;,
\label{par}
\end{equation}
we have obtained the constants $c_f=1.32$ and $c_g=-1.19$, and the powers 
$\alpha_f=5.18$ and $\alpha_g=5.14$ by fitting Eq.~(\ref{par}) the PQCD
results at large $\rho$ \cite{SLL2}. The values of the form factors at
zero recoil, $f_1(1)=1.32$ and $g_1(1)=-1.19$, are close to those
derived based on wave function overlap integrals \cite{XHG}. The
approximate equality of $f_1$ and $|g_1|$ is consistent with the
conclusion drawn from heavy quark effective theory (HQET) \cite{GG,FN}.

We emphasize that there is no conflict between PQCD and HQET, which has
been applied to heavy hadron decays successfully. If a hadron is indeed
very heavy, a small portion of its mass used for the energy release
involved in decay processes can guarantee the applicability of PQCD. That
is, the decay processes move into the perturbative regime quickly
once the velocity transfer is slightly greater than unity. In \cite{SLL2}
we have made a quantitative investigation to find out how far from
the zero recoil PQCD is applicable to $\Lambda_b$ baryon decays. Besides,
HQET gives the relations among various transition form factors, while
PQCD can be employed to calculate the behaviors of these form factors
near the high end of the velocity transfer. We have also performed the 
heavy quark expansion as constructing the heavy baryon wave functions. 
Hence, the two approaches in fact complement each other.

To extract the transverse polarization of a $\Lambda_b$ baryon, we
propose to measure a lepton asymmetry parameter in the laboratory frame. 
Choose a unit vector $\hat{e}$ perpendicular to the $\Lambda_b$ baryon 
momentum, whose direction is known if the interaction point of hadron
beams as well as the decay vertex of the $\Lambda_b$ baryon are observed. 
The lepton asymmetry parameter $\alpha(\hat{e})$ with respect to $\hat{e}$ 
is defined as 
\bea
\alpha(\hat{e}) &=& \frac{N_+ - N_-}{N_+ + N_-} \;,
\eea
where $N_+(N_-)$ is the number of leptons with positive (negative) values
of $\hat{\bf p}_l \cdot \hat{e}$. Since $\alpha(\hat{e})$ is invariant 
under a boost along the $\Lambda_b$ baryon momentum, we can 
simply calculate $\alpha(\hat{e})$ in the rest frame of the $\Lambda_b$ 
baryon, {\it i.e.}, in the above framework, and derive its relation to 
the $\Lambda_b$ baryon polarization.

Integrating over the variables $E'$ (or $\rho$) and $E_l$ in 
Eq.~(\ref{dr3}), we arrive at
\bea
\frac{d\Gamma}{d\Omega_l} &=& \frac{\Mc G_F^2}{16 \pi^4}
           |V_{cb}|^2 \; ( G_0 + G_1 \; \hat{\bf p}_l\cdot{\bf P} )\;,
\label{GO}
\eea
with the integrals
\bea
G_0 &=& \int d\rho dE_l \; F_0(E^\prime,E_l)\;,
\nonumber \\
G_1 &=& \int d\rho dE_l \; [ \Mc\sqrt{\rho^2-1}\cos\bar{\theta} \;
F_1(E^\prime,E_l) + E_l \;F_2(E^\prime,E_l) ] \;.
\label{G0G1}
\eea
It is straightforward to show, from Eq.~(\ref{GO}),
\bea
N_\pm &=& C_0 (G_0 \pm\frac{1}{2} G_1 \;{\bf P}\cdot \hat{e})  \;,
\eea
$C_0$ being a constant depending on the production rate of $\Lambda_b$ 
baryons. Therefore, we have
\bea
\alpha(\hat{e}) = \frac{G_1}{2 G_0} \;{\bf P}\cdot \hat{e}=
-0.31{\bf P}\cdot \hat{e}  \;,
\label{asyv}
\eea
where the numerical results of $G_0$ and $G_1$ from Eq.~(\ref{G0G1})
have been inserted. It implies that the lepton tends to be emitted in the 
direction opposite to the transverse polarization of the $\Lambda_b$ 
baryon and that the magnitude of the asymmetry parameter can not exceed 
0.31.

We have tested the sensitivity of our predictions to the variation of the
heavy baryon wave functions. The asymmetry parameters $\alpha(\hat{e})$
change only few percents for various choices of the baryon wave
functions. This observation is expected, since nonperturbative effects
from the wave functions cancel in the ratio of $(N_+-N_-)/(N_++N_-)$. For
a similar reason, this ratio does not depend on the normalizations of the
heavy baryon wave functions.  Therefore, Eq.~(\ref{asyv}) can be regarded
as being almost model-independent, and verified experimentally. Other 
potential corrections to the asymmetry parameter are under control and can 
be included systematically. For example, higher-order corrections to the
hard amplitudes are suppressed by $\alpha_s/\pi< 0.3$, which give 30\%
uncertainty at most. Higher-twist contributions are of order
${\bar \Lambda}/M_{\Lambda_b}\sim 0.1$ with the mass difference
${\bar\Lambda}=M_{\Lambda_b}-m_b$, $m_b$ being the $b$ quark mass.

The $\bar{\Lambda}_b$ baryon, produced together with the $\Lambda_b$ 
baryon in proton-anti-proton collisions, possesses a polarization 
opposite to that of the $\Lambda_b$ baryon, while the lepton from the 
$\bar{\Lambda}_b\to \bar{\Lambda}_c {\bar l}\nu$ decay tends to be 
produced along the transverse polarization of the $\bar{\Lambda}_b$ 
baryon. As a result, for colliding experiments at the Tevatron, it is 
possible to combine events of the $\Lambda_b$ and $\bar{\Lambda}_b$ 
baryons to double the statistics, and measure the lepton asymmetry 
parameter irrespective of whether the leptons are from $\Lambda_b$ or 
$\bar{\Lambda}_b$ baryon decays. With about 200 $\Lambda_b$ baryon
events available currently, the asymmetry parameter could be
measured to 10\% accuracy corresponding to a 30\% polarization. For
Run II of the Tevatron, it is likely to collect 20 times more events so 
that a polarization of a few percent could be determined. In addition, 
the $\Lambda_b$ baryon polarization can be measured through meson 
asymmetry parameter in nonleptonic decays. The application of the PQCD 
formalism to this case is under investigation.

\section{CONCLUSION}

In this paper we have studied the polarized $\Lambda_b$ baryon
semileptonic decay $\Lambda_b \to \Lambda_c l\bar{\nu}$ using PQCD
factorization theorem. We have proposed a boost invariant quantity
$\alpha(\hat{e})$ related to the transverse polarization
${\bf P} \cdot \hat{e}$ of a $\Lambda_b$ baryon, which can be measured
with available data from current and future experiments. This quantity is
defined as the number asymmetry of the leptons which are produced with
positive and negative components of momenta along $\hat{e}$. With respect
to the production plane of $\Lambda_b$ baryons at the Tevatron, it is the
asymmetry of the lepton numbers above and below this plane. Using the
transition form factors obtained in the PQCD approach, we have predicted 
$\alpha(\hat{e}) = -0.31 {\bf P} \cdot \hat{e}$. It indicates that the
lepton tends to be emitted in the direction opposite to the $\Lambda_b$ 
baryon transverse polarization and that the magnitude of the 
asymmetry parameter can not exceed 0.31. Our result is useful for 
extracting the transverse polarization of a $\Lambda_b$ baryon from its 
exclusive semileptonic decay.

\vskip 1.0cm
This work was supported by the National Science Council of the Republic of
China under Grant Nos. NSC-88-2112-M-001-041 and NSC-89-2112-M-006-004.

\vskip 3.0cm

\begin{table}
\begin{tabular}{ccccc}
$C$ & 0.2 & 0.3 & 0.4 & 0.5 \\
\hline
$\rho=1.2$  & 18\% & 49\% & 66\% & 78\% \\
$\rho=1.3$  & 22\% & 54\% & 71\% & 79\% \\
$\rho=1.4$  & 25\% & 58\% & 74\% & 82\% 
\end{tabular}
\caption{Percentages of the full contributions to the form factor
$f_1$, that arise from the region with $\alpha_s/\pi < C$.}
\end{table}
\end{document}